\documentclass[draft]{agujournal2019}
\usepackage{url} 
\usepackage{lineno}
\usepackage[inline]{trackchanges} 
\usepackage{soul}

\draftfalse
\journalname{Geophysical Research Letters}

\begin{document}




\title{The upper mantle of Europe and Western Asia: Exposing potentials and limits of regional full-waveform inversion}


\authors{Carl J.  Schiller\affil{1}, Scott D.  Keating\affil{1}, Sebastian Noe\affil{1}, Solvi Thrastarson\affil{1}, Dirk-Philip van Herwaarden\affil{1}, Christian Boehm\affil{1, 2}, Arthur J.  Rodgers\affil{3}, Pablo Barrera-Lopez\affil{1, 4}, Patrick Marty\affil{2} and Andreas Fichtner\affil{1}}

\affiliation{1}{Department of Earth and Planetary Sciences, ETH Zurich,  Zurich, Switzerland}
\affiliation{2}{Mondaic Ltd., Zurich, Switzerland}
\affiliation{3}{Geophysical Monitoring Program, Lawrence Livermore National Laboratory, Livermore, CA, USA}
\affiliation{4}{École et Observatoire des Sciences de la Terre, Université de Strasbourg, Strasbourg, France}

\correspondingauthor{Carl J. Schiller}{carl.schiller@eaps.ethz.ch}


\begin{keypoints}
\item Full-waveform inversion of Europe and Western Asia, increasing data volume by an order of magnitude compared to previous studies.
\item Enables the use of shorter-period full-waveform data to reduce uncertainties of earthquake moment tensor estimates.
\item Wavefield-adaptive meshes shift the bottleneck of regional full-waveform inversion from compute power back to data coverage.
\end{keypoints}


\begin{abstract}
We present EUWA310, a seismic velocity model obtained by full seismic waveform inversion (FWI) model of the upper mantle beneath Europe and Western Asia.  Inferred via 310 quasi-Newton updates from more than 260$\,$000 three-component seismograms at 18 s minimum period. It increases the number of iterations and the amount of data by an order of magnitude, compared to previous FWI studies in the region.  The data include recordings from two dense regional arrays: IberArray on the Iberian Peninsula and in Northern Africa, and AlpArray in the wider Alpine region.  The construction of EUWA310 became computationally tractable through the use of dynamic mini-batch optimization and the adoption of wavefield-adaptive spectral-element meshes from the global to regional scales.  We also demonstrate that EUWA310 may be used to reduce uncertainties of moment tensor estimates by inverting waveforms with decreasing minimum period until an optimal minimum period of 22 s, below which no additional improvements is achieved. This limit is connected to the termination of the FWI at 18 s period,  which did not result from limited computational resources, but from failure to explain data at even shorter periods.  Since other continental-scale arrays -- such as USArray or ChinaArray -- are comparable in density and coverage, this observation suggests that we may have reached the regime where data coverage instead of compute power is the main bottleneck in regional FWI.
\end{abstract}


\section*{Plain Language Summary}

In this paper, we are resolving for geological structures over most of the Eurasian continent as well as parts of Africa and the Arabian peninsula. We mostly use publicly available seismograms (ground motion recorded at seismic stations) measured in Europe, North Africa, the Arabian peninsula, Central Asia, India and Western China. We resolve the volume between different earthquake - station pairs with the full-waveform inversion (FWI) imaging method, resulting in a model of the seismic wave speed from the crust to the core-mantle boundary in our domain. In seismic wave speed we can see some known geological structures beneath mountain orogens, volcanoes or larger graben structures. The FWI implementation in this regional study additionally benefits from bounded adaptive meshes that further reduce computational costs. This study resulted in the highest resolved regional model of Europe and Western Asia to date, where we also assess the resolvability of earthquake sources in the region.


\section{Introduction}

The European continent and its tectonically active surroundings have been the subject of seismic tomography studies since the earliest days of this branch of science.  In the late 1970s, this region served as a testing ground for several pioneering developments,  including teleseismic body wave tomography \cite{Aki_Lee_1976,Aki_1977} and multi-mode Rayleigh wave inversion \cite{Nolet_1976,Nolet_1977}.  Since then, the comparatively densely instrumented continent has been illuminated with a wide range of methods, exploiting all possible seismic wave types \cite{Nolet_1990,Zielhuis_Nolet_1994,Spakman_1991,Spakman_1993,Kustowski_2008b,Peter_2008,Schaefer_2011,Schivardi_2011,Verbeke_2012,Legendre_2012}.

Two of the latest continental tomographic studies that considered the whole European continent include the full-waveform inversions (FWIs) of \citeA{Fichtner_2013} and \citeA{Zhu_2015}.  Combining numerical wave simulations and adjoint techniques \cite{Virieux_2009,Fichtner_book,Liu_2012}, FWI naturally enables the joint inversion of all body and surface wave phases,  eliminates the need for crustal corrections,  and permits iterative updates that account for nonlinearity in the inversion. Resulting FWI models often explain phases and amplitudes of previously unseen waveform recordings \cite{Tape_2010,Bozdag_2016,Thrastarson_2024}. This ability of FWI models to generalize is the foundation of efforts to improve moment tensor estimates by inverting complete seismograms at increasingly shorter periods and to avoid biases induced by the use of 1-D Earth models \cite{Liu_2004,Hingee_2011,Hejrani_2017,Sawade_2022,Doody_2025}.

More than 10 years after their construction, the last two FWI models of Europe \cite{Fichtner_2013,Zhu_2015} appear strongly limited by the availability of both seismic data and computational resources.  They incorporated a few tens of thousands of recordings from several hundred stations and performed few tens of iterations to improve the models; compared to several hundred iterations in recent regional- to global-scale FWIs \cite{Rodgers_2024,Thrastarson_2024}. 

During the past decade, the seismic tomography landscape has advanced substantially.  Waveform data from numerous additional permanent and temporary deployments have become available.  The latter include IberArray with $\sim$200 stations on the Iberian Peninsula and in Morocco \cite{IberArray_2007,Diaz_2009}, as well as AlpArray with $\sim$400 stations in the broader Alpine region \cite{AlpArray_2015,Hetenyi_2018}.  Simultaneously,  the incorporation of significantly larger data volumes has become possible thanks to the combined effect of steadily increasing computational power and methodological developments that reduce the computational requirements per assimilated seismogram. The latter include the translation of stochastic gradient methods from machine learning to seismic tomography \cite{vanLeeuwen_2013b,vanHerwaarden_2020},  and the development of numerical meshes with less elements that exploit wavefield symmetries \cite{Leng_2019,vanDriel_2020,Thrastarson_2020}.

In the light of these advances,  the key objectives of this work are as follows: (i) Produce a state-of-the-art FWI model of Europe and Western Asia by incorporating recent waveform data and exploiting wavefield-adaptive meshes that permit a larger number of iterative updates. (ii) Quantify the frequency-dependent ability of the resulting model to explain previously unseen data and to reduce uncertainties in moment tensor inversions. (iii) Assess the current limitations of continental-scale FWI and the implications for future FWI research.

\section{Data}\label{S:Data}

\begin{figure}
    \centering
    \hspace*{-1.8cm} 
    \includegraphics[width=1.0\linewidth]{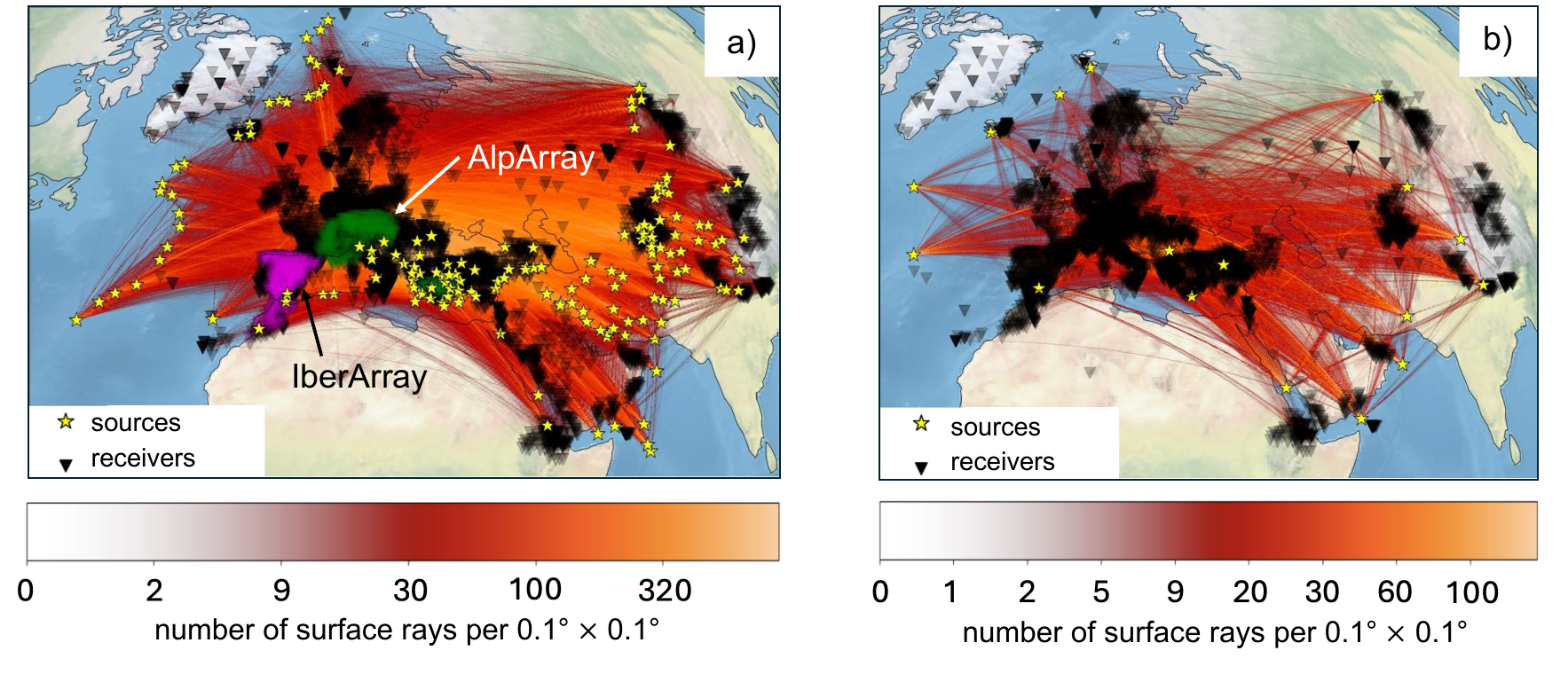}
    \caption{Data summary in the form of surface ray density plots.  a) The inversion dataset includes 163 sources (earthquakes, yellow stars) and 7$\,$431 three-component stations (black triangles). Since not all sources were recorded by all stations, the total number of unique source-receiver pairs is 261$\,$085.  The color bar indicates the number of surface rays within a 0.1$^\circ \times$ 0.1$^\circ$ area.  b) Coverage of the validation dataset, comprising 17 sources not used in the inversion. We chose their locations to provide similar azimuthal coverage as the inversion dataset. }
    \label{raydensity}
\end{figure}

We chose the extent of the study area to provide dense source-receiver coverage of Europe and Western Asia. In addition to the European continental plate, the study area encompasses the northern part of the mid-Atlantic ridge, Greenland, Iceland, Jan-Mayen, the Middle East,  the Caucasus, West-Central Asia, the Hindu Kush, and parts of the Himalayas.  Using the FDSN web services \cite{romanowicz1986toward}, we downloaded seismic waveforms for earthquakes in the magnitude range from $M_w$ 5.2 - 6.7,  as listed in the GCMT Catalog \cite{ekstrom2012global}. This range represents an excellent balance between events that are large enough to provide a useful signal-to-noise ratio for most stations, while being small enough to avoid finite-source effects \cite{vallee2013source}. \\[5pt]
Our dataset includes 163 earthquakes, recorded at 7$\,$431 stations. In total, there are 261$\,$085 unique source-receiver pairs,  with epicentral distances up to 95$^\circ$.  In addition to numerous permanent stations that were not yet available for the previous European FWI studies of \citeA{Zhu_2012} and \citeA{Fichtner_2013}, the dataset includes recordings from two recent large dense arrays: (i) $\sim$200 IberArray stations on the Iberian Peninsula and in Morocco, with an average station spacing on the order of 100 km \cite{IberArray_2007,Diaz_2009}, and (ii) $\sim$400 stations of AlpArray in the broader Alpine region, with an average station spacing of around 50 km \cite{AlpArray_2015,Hetenyi_2018}. The data coverage is visualized in Fig. \ref{raydensity}a.  We select a subset of 17 well-distributed events, shown in Fig.  \ref{raydensity}b,  as validation dataset, which is not used in the inversion to update the model. Instead, it serves to assess the model's ability to generalize to unseen data and to avoid over-fitting by terminating the inversion when the validation misfit increases over several subsequent iterations.
%
%
For the inversion, we use recordings that begin 300~s prior to the GCMT origin time and last for 3$\,$900 s.  The first 300 s are needed to estimate noise levels, and the following 3$\,$600 s roughly correspond to the longest surface-wave travel time for the largest epicentral distance in the 18 - 50 s period band. Since our initial model,  the second-generation Collaborative Seismic Earth Model, CSEM2, \cite{noe2024collaborative}, largely explains seismic waveform data globally at a minimum period of 50 s,  we consider a maximum period of 50 s, aiming to reach 18~s for the final model.

\section{Methodology}

\subsection{Forward and adjoint modeling}

For the seismic wavefield simulations, we employ a variant of the spectral-element method that uses wavefield-adaptive meshes \cite{vanDriel_2020}, which harnesses approximate wavefield symmetries and wavelength anisotropy.  In azimuthal directions (parallel to the wavefront), the wavelength is typically much larger than in radial directions (perpendicular to the wavefront). In smooth media, such as the Earth at regional to global scales,  where the seismic wavefield is dominated by transmission instead of scattering,  this allows us to coarsen the computational mesh in azimuthal directions.  Consequently, the number of grid points, and with it the computational requirements for 3-D simulations, can be reduced substantially.  The extent to which the meshes can be coarsened azimuthally depends on the complexity of the medium and the minimum period. It needs to be adjusted as the inversion progresses by comparing to solutions for conventional 3-D meshes of the cubed-sphere type \cite{ronchi1996cubed, komatitsch1999introduction}.  Gradients corresponding to this forward problem solution are calculated through the adjoint method on the same wavefield-adaptive mesh. These gradients are exact to numerical precision: the adjoint problem on the wavefield-adaptive mesh is the correct complementation to the forward solution and not an approximation. Hence, the cost savings are preserved through the adjoint propagation without additional approximation \cite{Thrastarson_2020}.\\[5pt]
Our main methodological development is the extension of the wavefield-adaptive mesh approach from the global scale,  where it was initially developed and applied \cite{thrastarson2022data,Thrastarson_2024}, to the regional scale.  This involves a modification of mesh generation routines and the incorporation of absorbing boundaries.  The supplemenary Fig.  \textcolor{red}{S1} illustrates the efficiency of the wavefield-adaptive meshes for the final model, presented in detail in Sec. \ref{S:Model}. It shows seismograms at 18 s minimum period computed with a wavefield-adaptive mesh composed of 1$\,$280$\,$000 elements. For the reference seismograms, we used a standard cubed-sphere-type mesh without azimuthal coarsening, containing 5$\,$875$\,$000 elements. The radial grid spacing of both meshes is identical.  As expected, waveform differences increase slightly with increasing epicentral distance and recording time, but remain significantly smaller than the achievable differences between simulated and observed seismograms (see Fig.  \ref{fig:waveform_comparison} for examples). At longer periods, the agreement between the seismograms for the two types of meshes generally increases, meaning that the comparison at the shortest period of 18 s is conservative.\\[5pt]
To further reduce computational cost, we implement dynamic mini-batch optimization \cite{bottou2010proceedings, vanLeeuwen_2013b, boehm2018time, vanHerwaarden_2020}, which exploits data redundancies resulting from similar epicenter locations.  Mini-batches contain quasi-random subsets of sources that are large enough to approximate the misfit gradient of the entire dataset sufficiently well to ensure rapid convergence.  The benefit of mini-batches is to effectively break the linear scaling of compute cost with the total number of sources \cite{vanHerwaarden_2020}. \\[5pt]
In this work, we combine wavefield-adapted meshes with dynamic mini-batch optimization, similar to \citeA{thrastarson2022data}, who achieved a cost per iteration of $\sim$0.6\,\%, compared to standard FWI using cubed-sphere-type meshes and all events in each iteration. We employ the spectral-element solver Salvus \cite{afanasiev2019modular} to perform forward and adjoint simulations. The implementation includes ocean loading, viscoelastic attenuation \cite{robertsson1994viscoelastic}, as well as ellipticity and topography of the Earth.

\subsection{Misfit quantification and validation}

To quantify discrepancies between observed and simulated waveforms, we employ time-frequency phase misfits \cite{fichtner2008theoretical}, which measure frequency-dependent phase shifts in time-frequency windows where waveform differences are small enough to prevent cycle skipping. At the same time, they are invariant with respect to amplitude scaling, which is beneficial when earthquake magnitudes are potentially inaccurate.  As a measure to balance dense station clusters against geographically isolated stations, we apply station weights, following the scheme of \citeA{ruan2019balancing}.  It has the effect of model-space preconditioning that accelerates convergence \cite{van2023full}. \\[5pt]
We calculate validation misfits for events in Fig.  \ref{raydensity}b to assess the model's ability to explain data not used in the inversion and to prevent over-fitting.  To avoid a positive bias that could be introduced by limiting phase misfit measurements to time windows without cycle skips, we adopt the suggestion of \cite{Tape_2010} and use the $L_2$-waveform misfit of complete seismograms as a conservative and less biased validation misfit.  Consequently, both phase and amplitude discrepancies between observed and computed seismograms are included in the validation. 

\subsection{Multiscale inversion}

A central requirement of iterative gradient-based FWI is the avoidance of cycle skipping, which may cause convergence towards meaningless local minima.  In addition to making measurements in carefully selected windows, it has two components: (i) the choice of a suitable initial model, and (ii) the progression from long to short periods \cite{bunks1995multiscale, fichtner2013multiscale}.  Our initial model is the second-generation Collaborative Seismic Earth Model (CSEM2) \cite{noe2024collaborative}, constructed to fit seismic waveforms with a minimum period of $\sim$50 s at global scale.  Borrowing the CSEM2 parametrization, we describe our model in terms of the wave speeds of vertically and horizontally polarized S waves ($V_{SH}$ and $V_{SV}$), the wave speeds of vertically and horizontally propagating P waves ($V_{PH}$ and $V_{PV}$), mass density $\rho$,  P- and S-wave attenuation ($Q_{P}$ and $Q_{S}$),  and the dimensionless parameter $\eta$ that captures variations in P-wave speed with incidence angle \cite{babuska1991seismic}.  Since 3-D variations in attenuation and P-wave anisotropy are difficult to resolve at continental scale,  we only update the isotropic P-wave speed $V_{P}$,  set $\eta = 1$, and keep attenuation equal to the 1-D profile in CSEM2, which is equivalent to the attenuation structure of PREM \cite{Dziewonski_Anderson_1981}.  We partition the inversion into multiple stages, progressing from the longest minimum period of 50 s to the shortest period of 18 s. \\[5pt]
The complete model is represented on a spectral-element master mesh of the cubed-sphere type.  In each iteration, the workflow begins with the selection of a random mini-batch of sources, following the procedure described in \citeA{vanHerwaarden_2020}.  To enable fast forward and adjoint simulations, we generate a unique wavefield-adaptive simulation mesh for each of the sources and interpolate the current model from the master mesh.  After the adjoint simulation,  we project the source-wise gradients back onto the master mesh and sum over all sources to obtain the raw total gradient.  We prevent the appearance of small-scale heterogeneities that are unlikely to be resolved by applying a Gaussian smoothing filter with a width of $\sim$1 and $\sim$0.5 minimum S-wavelengths in horizontal and vertical directions, respectively. This produces a preconditioned gradient that drives an L-BFGS trust-region optimization \cite{Nocedal_1999,vanHerwaarden_2020}.

\section{Model presentation}\label{S:Model}

\subsection{Iterative model evolution and final model}

\begin{figure}
    \centering
\hspace*{-2.5cm} 
    \includegraphics[width=1.0\textwidth]{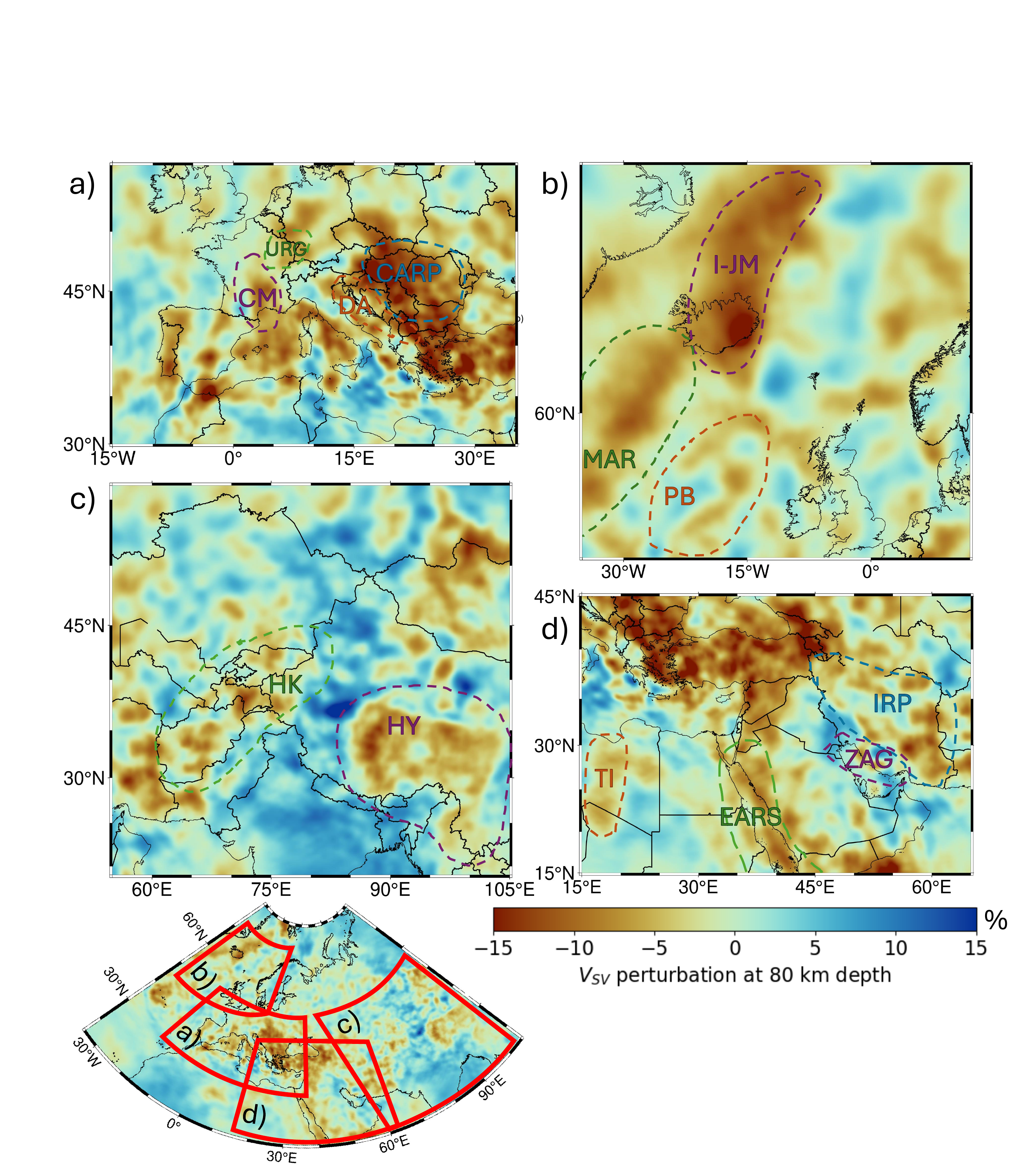}
    \caption{Variations of $V_{SV}$ at 80 km depth, including several close-ups. Some known geological features can be identified in the close-ups. a) URG: Upper Rhine Graben, CM: Central Massif, CARP: Carpathian Mountains, DA: Dinaric Alps. b) I-JM: Iceland-Jan Mayen plume system, MAR: Mid-Atlantic Ridge, PB: Porcupine Bank and Rockall Through. c) HK: Hindu Kush mountains, HY: Himalayas. d) TI: Tibesti mountains, IRP: Iranian Plateau, ZAG: Zagros mountains, EARS: East African Rift System. }
    \label{fig:EUWA_80km}
\end{figure}

In total, we performed 310 iterations, leading from the initial model CSEM2 to the final model of Europe and Western Asia, EUWA310.  A detailed listing of misfit reductions per iteration stage (period band) can be found in Supplementary table \textcolor{red}{T1}. Supplementary Fig. \textcolor{red}{S2} shows that the boundaries of wave speed anomalies generally sharpen as we reach shorter periods,  and allows the identification of more features with scale lengths on the order of 100 km. \\[5pt]
Fig. \ref{fig:EUWA_80km} visualizes $V_{SV}$ variations at 80 km depth in EUWA310, including close-ups of several subregions. These exhibit the tomographic expression of numerous well-known geologic structures,  including the Massif Central, the Apennines, the Carpathians, the Dinarides,  and the Upper Rhine Graben.  In the south-eastern part of the model, we can distinguish the East African Rift System, the Tibesti-, Zagros- and Hindu Kush mountains,  as well as the Iranian plateau.  Additional depth slices through both the $V_{SV}$ and $V_{SH}$ models can be found in the supplementary Fig.  \textcolor{red}{S3} and Fig. \textcolor{red}{S4}, respectively.
%

\subsection{Waveform fit}

\begin{figure}[h!]
\hspace*{-1.6cm}
    \centering
    \includegraphics[width=0.9\linewidth]{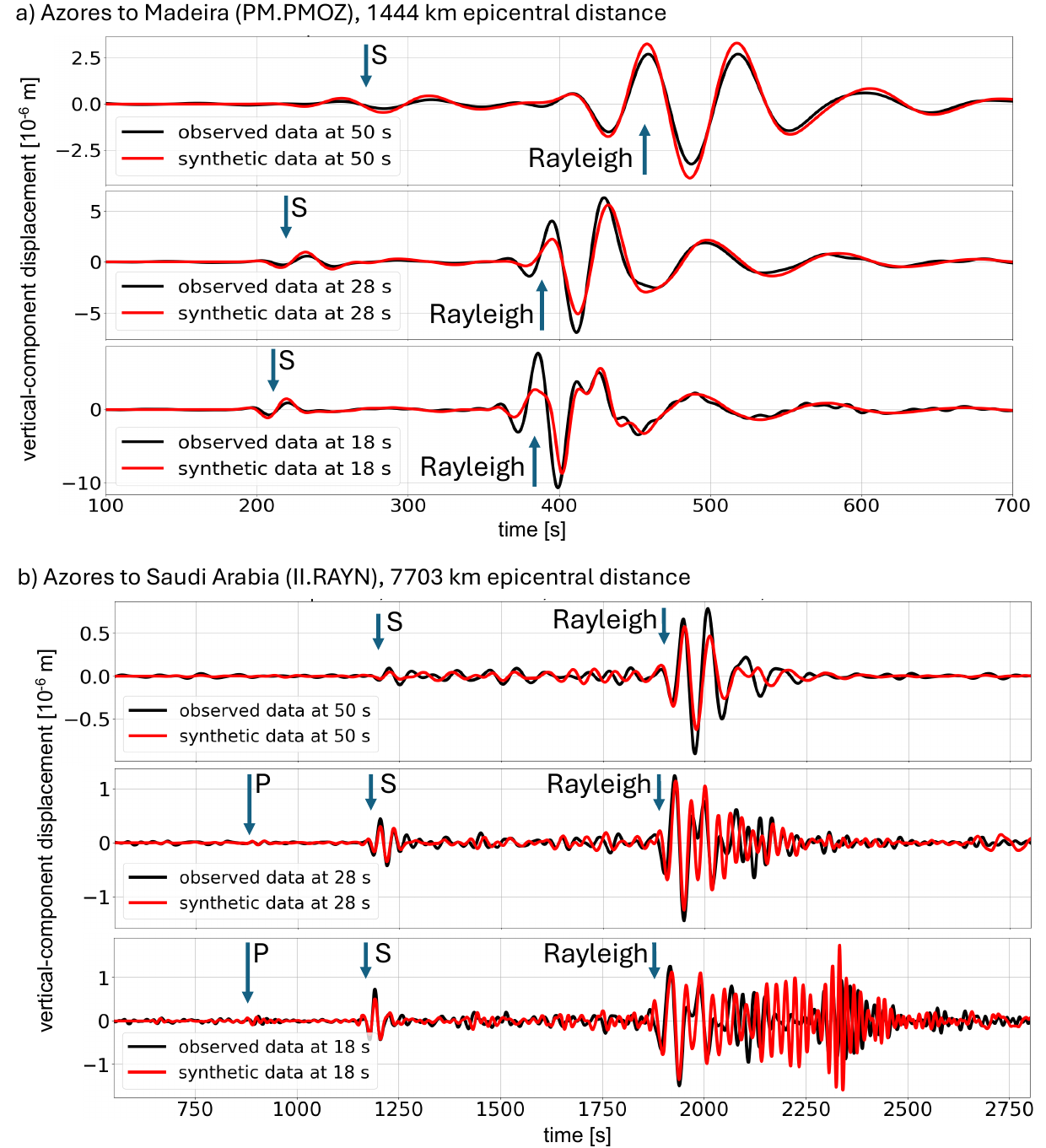}
    \caption{Comparison of observed (black) and simulated (red) vertical-component waveforms fit in the period bands from 50 - 90 s,  28 - 90 s and 18 - 90 s for an earthquake that occurred near the Azores. a) At station PM.PMOZ (Madeira,  1$\,$444 km epicentral distance) both body and surface waves can be matched at periods as low as 18 s, although with increasing misfit towards shorter periods. b) At station II.RAYN (Saudi Arabia, 7$\,$703 km epicentral distance),  waveform fit is acceptable only for body and fundamental-mode Rayleigh waves. Later-arriving surface waves cannot be explained at the shortest period of 18 s.}
    \label{fig:waveform_comparison}
\end{figure}

Although the overall misfit decreases markedly during the iterative inversion, misfits for individual source-receiver pairs depend strongly on epicentral distance and minimum period.  This has implications for the predictive power of the model and its use, e.g., in earthquake source inversion. To exemplify this dependence, we compare observed and computed waveforms in different period bands of the 2016 Azores earthquake for two cases that are representative for the entire dataset: (i) a shorter epicentral distance of 1$\,$444 km to station PM.PMOZ on Madeira in Fig. \ref{fig:waveform_comparison}a,  and (ii) a larger epicentral distance of 7$\,$073 km from the Azores to station II.RAYN in Saudi Arabia in Fig.  \ref{fig:waveform_comparison}b.\\[5pt]
At shorter distances, complete waveforms can be matched well at 18 s period.  In contrast, at longer distances, only body wave arrivals -- and occasionally the fundamental-mode Rayleigh wave -- can be explained at 18 s period.  Especially the observed surface wave trains following the fundamental model below $\sim$25 s period often become too complex to be matched by the simulations.  In addition to the vertical-component Rayleigh waves in Fig. \ref{fig:waveform_comparison}, also horizontal-component Love waves at large epicentral distances hardly match at 18 s period. The simulated Love waves tend to be too fast, indicating that some low-wave-speed anomalies cannot be reconstructed with the available data coverage.  This overall trend is also reflected in the relative misfit reductions in the supplementary Table \textcolor{red}{T1},  which are generally more significant at longer periods. Since our inability to match surface waves dominates the total misfit below 18 s period, we did not continue to shorter periods and terminated the inversion.
%

\section{Moment tensor inversion}

\begin{figure}
    \centering
    \includegraphics[width=1.0\linewidth]{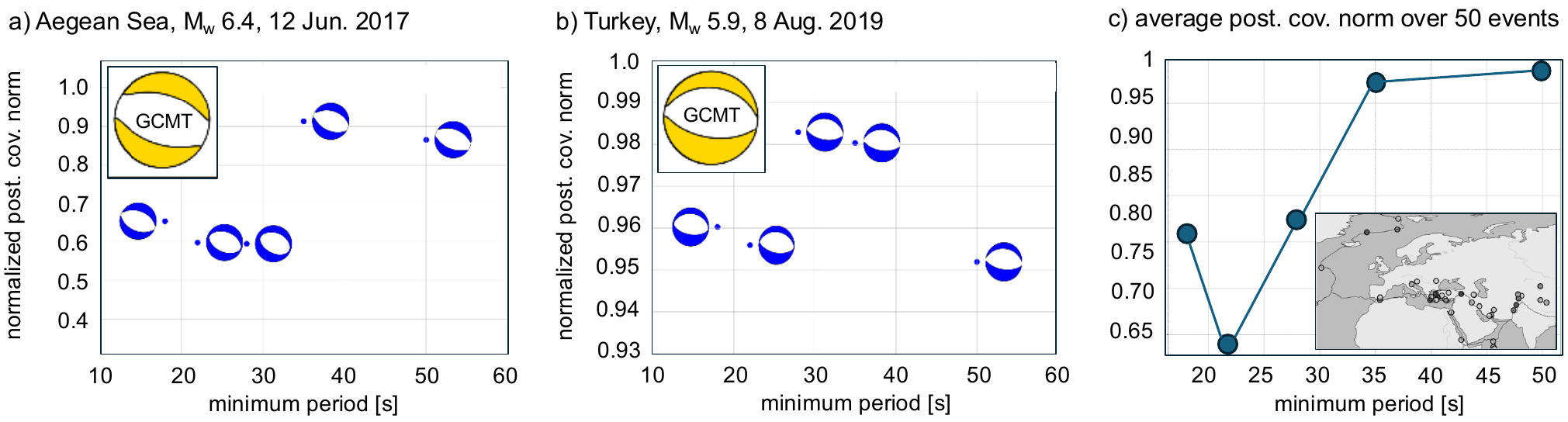}
    \caption{Norm of the posterior covariance matrix (normalized) as a function of the minimum period used for moment tensor inversion. a) Example for an $M_w$ 6.4 earthquake in the Aegean Sea. Reducing the minimum period from 50 s to 22 s reduces the posterior covariance by $\sim$30 \%.  A further reduction in minimum period does not lead to further improvements. b) Example of an $M_w$ 5.9 event in Turkey.  Reducing the minimum period in the moment tensor inversion has no significant effect on the posterior covariance.  c) Average normalised posterior covariance norm for 50 earthquakes within the study region, the locations of which are shown in the inset. }
    \label{fig:mt_summary}
\end{figure}

In addition to incorporating large data volumes from recent dense arrays at the shortest possible period, a main objective of this work is the improvement of earthquake moment tensor inversion. As a baseline for comparisons, we use the GCMT catalog \cite{ekstrom2012global}, which estimates moment tensors from body and surface waves at periods above 50 s, and employs the 1-D PREM model \cite{Dziewonski_Anderson_1981} with 3-D path corrections.  This approximation may result in over-estimations of seismic moments and source depths,  and under-estimation of double-couple components \cite{abercrombie2001earthquake, patton2002causes, hjorleifsdottir2010effects, Sawade_2022}.  Using 3-D models for source inversion can overcome these issues and lead to improved waveform fits \cite{liu2004spectral, Hingee_2011,lee2011rapid, covellone2012quantitative, bozdaug2016global, Hejrani_2017,Sawade_2022}. \\[5pt]
Deviating from these earlier studies, our interest is less in changes of the moment tensor solution caused by the use of 3-D models, but in the reduction of uncertainties that we hope to achieve by incorporating more waveform observations at shorter periods.  For this,  we partition the data into the period bands of 50 - 90 s,  35 - 90 s, 28 - 90 s,  22 - 90 s and 18 - 90 s and conservatively select measurement windows where the phase difference is smaller than a quarter of the minimum period to avoid cycle skipping. We parametrize moment tensors in terms of the elementary strike-slip tensors \cite{kikuchi1991inversion} and exclude the isotropic contribution based on the prior knowledge that the events considered are tectonic strike-slip earthquakes.  With this, we can formulate the inversion as a well-posed linear least-squares problem that does not require regularisation and permits the explicit calculation of the posterior covariance matrix.\\[5pt]
Panels a) and b) in Fig. \ref{fig:mt_summary} display two representative moment tensor inversion examples; one from the Aegean Sea, the other from Turkey. For the Aegean event, the ability of the FWI model to incorporate shorter-period waveforms has a significant effect on the posterior covariance. Reducing the minimum period from 50 s to 22 s, reduces the posterior covariance norm by around 30 \%. Further reducing the minimum period to 18 s, does not yield additional improvements. In contrast, the Turkish event shows no significant improvement of the posterior covariance with decreasing period. In both cases, the orientation of the estimated moment tensors deviates only slightly from the GCMT solution, shown in yellow. This behaviour is characteristic for moment tensor inversions of a larger ensemble of 50 earthquakes,  summarized in Fig. \ref{fig:mt_summary}c. Moment Tensor inversions with our EUWA310 model show optimal variance reductions for a minimum period of 22 s, representing the shortest period where the model provides reliable waveform fits across the domain, with a posterior covariance reduction of around 40 \% relative to inversions at 50 s minimum period. For shorter periods, posterior covariances increase, on average.
%

\section{Discussion}

%



\subsection{Computational cost}\label{SS:Cost}

A central element of this work is the reduction in computational cost through the combination of wavefield-adaptive meshes and dynamic mini-batches.  The contribution of wavefield-adaptive meshes, visualized in supplementary Fig. \textcolor{red}{S5}, is to reduce the number of finite elements relative to a standard cubed-sphere-type mesh.  Averaged over the different stages of the inversion,  a wavefield-adaptive mesh requires $\sim$5 times less elements, which translates almost one-to-one into a reduced number of required GPU hours. \\[5pt]
The contribution of the mini-batch approach in our specific case is to decrease the per-iteration cost by about an order of magnitude relative to a full-batch strategy,  but the corresponding increase in convergence rate cannot be easily determined. Synthetic inversions suggest that it is around a factor of 4 \cite{vanHerwaarden_2020}. In combination, wavefield-adaptive meshes and dynamic mini-batches likely reduced the computational requirements by a factor of around 20,  thereby transforming such an inversion from being out-of-scale to being feasible with reasonable effort. Hence, for the complete inversion we used $\sim$400$\,$000 GPU hours on the Cray XC50 cluster \emph{Piz Daint}, operated by the Swiss Supercomputing Center (CSCS), which would have exceeded 2 million GPU hours if we were not to use either cost saving approach.

\subsection{Model comparison}\label{SS:Comparison}

\begin{figure}
    \centering
    \includegraphics[width=1.0\linewidth]{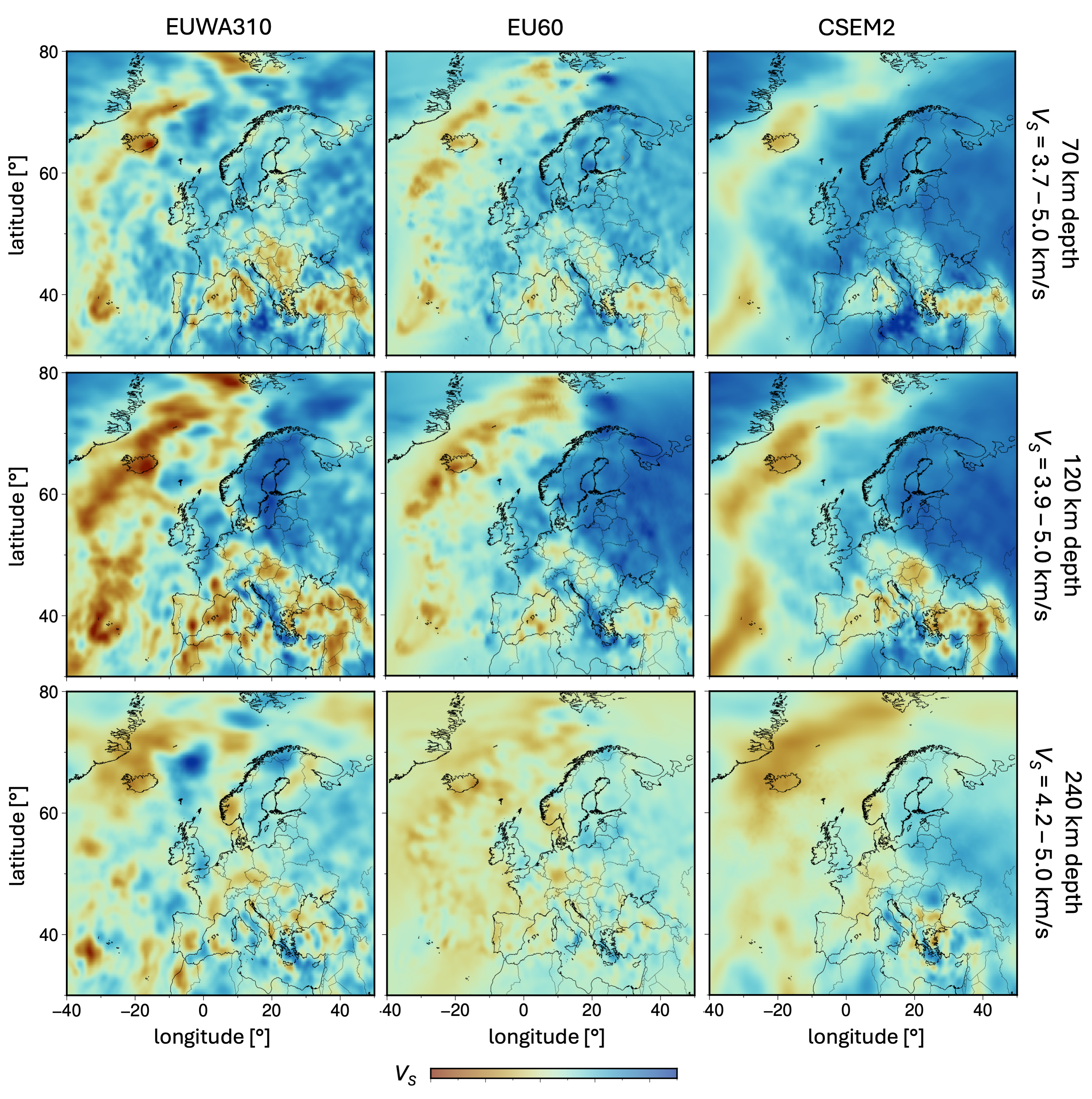}
    \caption{Comparison of absolute isotropic shear wave speed $V_S = \sqrt{(2V^2_{SV} + V^2_{SH})/3}$ \cite{babuska1991seismic} for three FWI models of Europe and Western Asia: EUWA310 (left), EU60 \citeA{Zhu_2015} (center), and CSEM2 \cite{noe2024collaborative} (right), which largely identical in this region to the continental-scale model of \citeA{fichtner2013multiscale}.  Shown are horizontal slices at 70 km (top), 120 km (middle) and 240 km (bottom) depth. The data range of the wave speed structure is shown explicitly for each depth.}
    \label{fig:compare_models}
\end{figure}

Quantitative comparisons of tomographic models are challenging due to (i) numerous subjective choices made during their construction \cite{Fichtner_2025}, (ii) different data and processing schemes, and (iii) the absence of a computationally tractable method for comprehensive uncertainty quantification in FWI that does not itself suffer from subjective choices and simplifications.  Here, we limit ourselves to a qualitative comparison for the purpose of demonstrating the geologic plausibility of our model and highlighting the effects of including new data, specifically from IberArray and AlpArray.\\[5pt]
Fig.  \ref{fig:compare_models} juxtaposes EUWA310 to the FWI model EU60 \citeA{Zhu_2015} and our initial model CSEM2 \cite{noe2024collaborative}. Within central Europe, the latter is largely identical to the regional-scale FWI model of \citeA{fichtner2013multiscale}.  While having used a similar methodology, both EU60 and the European part of CSEM2 are constrained by around an order of magnitude less data than EUWA310.  Furthermore, they have been constructed with longer minimum periods; 25 s for EU60 and 50 s for CSEM2.\\[5pt]
As expected, features with length scales above $\sim$500 km agree well at depths above $\sim$150 km, where coverage is dominated by surface waves. At greater depths, where body wave coverage becomes increasingly important, the models start to differ more significantly. Still, major features, such as the subducting lithospheric slabs of the wider Alpine orogen, are generally present. The effect of including more data at shorter periods and performing more iterations can be seen most clearly in both the sharpness and the strength of the heterogeneities, e.g.,  along the mid-Atlantic ridge.  Most clearly visible in Fig. \ref{fig:EUWA_80km},  numerous geologically plausible structures with $\sim$100 km scale length appear in central Europe, mostly likely as a result of including AlpArray and IberArray data. %

\subsection{Moment tensor inversion}\label{SS:moment_tensor}

The posterior covariances of the moment tensor inversions exhibit a pronounced period-dependence. While we observed covariances decrease on average as the minimum period decreases from 50 s to 22 s, any further reduction of the minimum period does not lead to additional improvements. This behaviour is the result of three main factors: (i) At longer periods, waveforms are simpler, and they can be fit by a larger variety of moment tensors. The larger effective nullspace is reflected in a higher posterior covariance in Fig. \ref{fig:mt_summary}c. (ii) Including shorter-period data, adds information, and thereby shrinks both the effective nullspace and its expression via the posterior covariance. This is the effect that we hope to exploit when using 3-D FWI models for moment tensor inversion, but is not reflected within every event in the dataset (iii) When the minimum period becomes too short,  the model fails to predict waveforms sufficiently well. Consequently, the number of time windows with an acceptable match between observed and simulated waveforms decreases, thereby effectively decreasing the amount of exploitable information. \citeA{chiang2026improved} recently showed that in general 3-D FWI models provide similar waveforms fits across distances and near-zero time-shifts with little variations across distances. 1D models perform poorer with worse variance reduction and more scattered time-shifts.

\subsection{Potentials and limits of current regional-scale full-waveform inversion}\label{SS:potentials}

Methodological improvements, such as wavefield-adaptive meshes and stochastic mini-batch optimisation, combined with increasing computational power, enable us today to assimilate large seismic waveform datasets collected at regional to global scales.  Just a few years ago,  FWIs like the one presented here may have terminated at long periods or after a small number of iterations due to insufficient computational resources.  Instead,  they may now stop because the inversion fails to explain shorter-period data, and specifically highly dispersed surface wave trains.\\[5pt]
This inability to converge towards a model that explains these data is the result of locking into a local minimum, from which a gradient-based descent method cannot escape.  Decreasing the minimum period by a factor of $x$, requires on the order of $x$ times more model parameters in each spatial direction, i.e., $x^3$ times more model parameters in total for a 3D problem.  However, constraining these additional parameters requires a corresponding increase in geographic data coverage.  In the inversion presented here, the lack of coverage terminates the iteration before computational limitations become a concern.\\[5pt]
While this statement is strictly valid only for our specific example, other continental-scale arrays, such as USArray (\url{http://www.iris.edu/hq/}) or ChinArray (\url{http://www.chinarraydmc.cn}), are similar in design. Station spacings are on the order of 100 km, and azimuthal illumination by sufficiently large earthquakes is sparse.  In summary, this suggests that we may be back, or are at least moving towards, a situation similar to the earliest days of seismic tomography where data and not compute power were the bottleneck.

\section{Conclusions}

The combination of dynamic mini-batch optimization and wavefield-adapted spectral-element meshes can reduce the computational cost of regional FWI by more than an order of magnitude.  In the case of Europe and Western Asia, it enables the assimilation of a large seismic waveform dataset -- including IberArray and AlpArray recordings -- around 10 times the size of datasets used in previous FWI studies of that region.  The resulting upper-mantle model, EUWA310, can benefit earthquake moment tensor inversion by allowing the incorporation of shorter-period waveforms down to a minimum period of $\sim$22 s. Although waveforms with a minimum period of 18 s entered the construction of EUWA310, periods lower than 22 s do not further reduce moment tensor uncertainties. The benefit of additional information from shorter periods is offset by a reduced number of time windows where observations and simulations match sufficiently well.  This limitation reflects the bottleneck in the construction of EUWA310, which is not a lack of computational power but insufficient data coverage that cannot prevent becoming trapped in local minima at periods below 18 s.  Since other continental-scale arrays are comparable in density and azimuthal illumination, this suggests that FWI may have reached a regime where computational resources are not the main limitation anymore.

\section{Open Research}

\textcolor{red}{To be written.}





\acknowledgments

The authors would like to thank Solvi Thrastarson, Dirk-Philipp van Herwaarden and Vaclav Hapla for their work on the software stack consisting of LASIF 2.0 \cite{thrastarson2021lasif}, MultiMesh \cite{thrastarson_2021_4564523}, Inversionson \cite{thrastarson2021inversionson} and Optson (van Herwaarden \& Hapla, 2023), which also use ObsPy \cite{beyreuther2010obspy, megies2011obspy, krischer2015obspy}, Pyasdf \cite{krischer2016adaptable}, SciPy \cite{virtanen2020scipy} and NumPy \cite{harris2020array} underneath. All computations have been done with Python \cite{rossum1995python} or C++ \cite{stroustrup2013c++}. More specifically with PETSc \cite{balay2019petsc} and Salvus \cite{afanasiev2019modular}. Depth slices and maps were created with Generic Mapping Tools \cite{wessel2019generic} using color maps from \citeA{crameri2018scientific}.

\textcolor{red}{To be written.}



\bibliography{agusample} 

\end{document}


%



\title{SUPPLEMENT TO\\[5pt]The upper mantle of Europe and Western Asia: Exposing potentials and limits of regional full-waveform inversion}


\authors{Carl J.  Schiller\affil{1}, Sebastian Noe\affil{1}, Dirk-Philip van Herwaarden\affil{1}, Christian Boehm\affil{1, 2}, Soelvi Thrastarson\affil{1}, Scott D.  Keating\affil{1}, Arthur J.  Rodgers\affil{3}, Pablo Barrera-Lopez\affil{1, 4}, Patrick Marty\affil{1} and Andreas Fichtner\affil{1}}

\affiliation{1}{Department of Earth and Planetary Sciences, ETH Zurich,  Zurich, Switzerland}
\affiliation{2}{Mondaic Ltd., Zurich, Switzerland}
\affiliation{3}{Geophysical Monitoring Program, Lawrence Livermore National Laboratory, Livermore, CA, USA}
\affiliation{4}{École et Observatoire des Sciences de la Terre, Université de Strasbourg, Strasbourg, France}

\correspondingauthor{Carl J. Schiller}{carl.schiller@eaps.ethz.ch}

\newpage


\section{Comparison of wavefield simulations}

%
\begin{figure}
    \centering
    \includegraphics[width=1.0\linewidth]{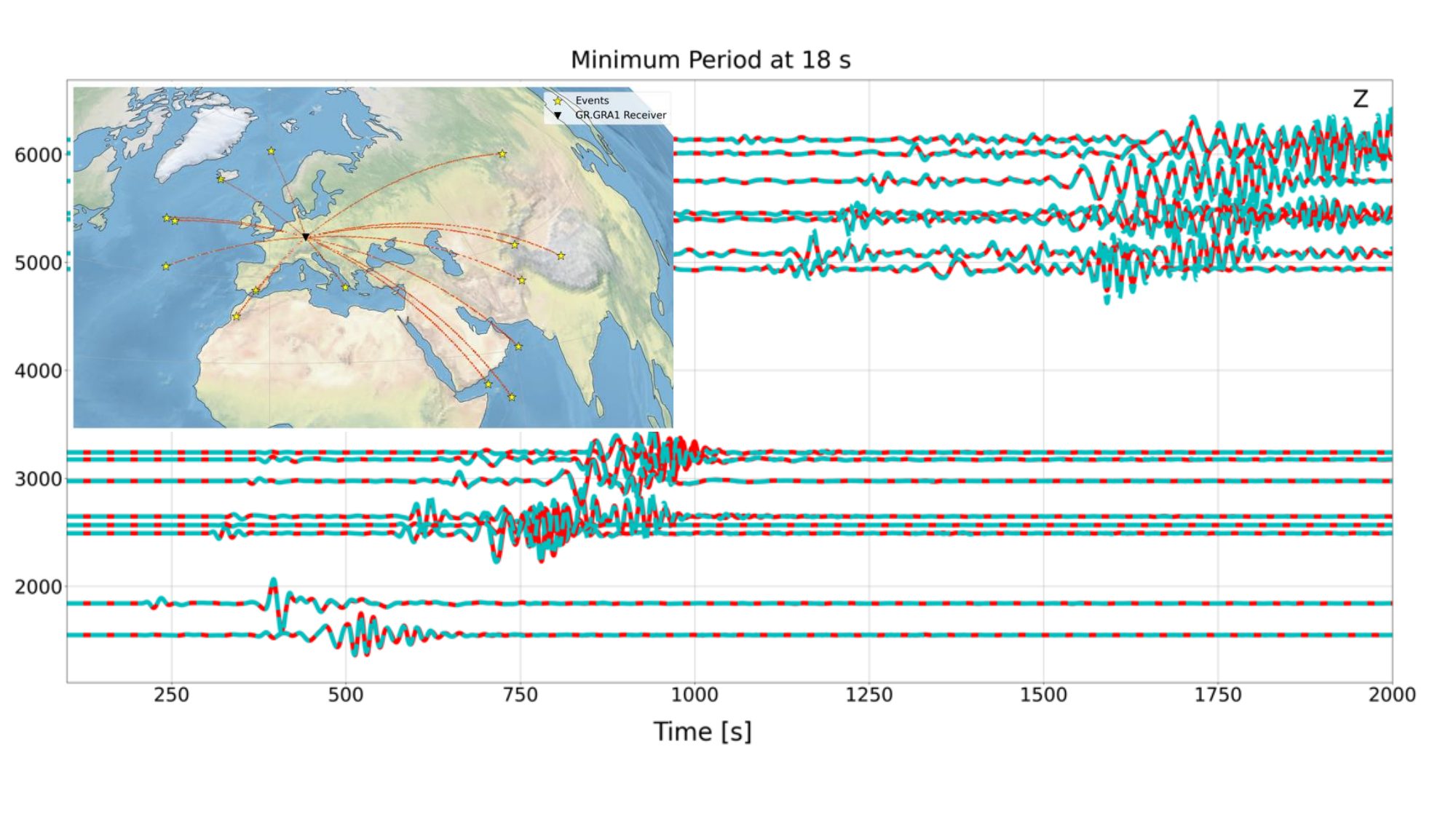}
    \caption{ Comparison of several vertical-component seismograms computed with a cubed-sphere type mesh (cyan) and a wavefield-adapted mesh (red). Station GR.GRA1, for which we computed the seismograms, is located in the center of the domain, and the 15 example events are at near-maximum distance close to the domain boundary.  }
    \label{fig:mesh-type-comparison}
\end{figure}
%


\section{Iteration development}

%
\begin{table}
 \caption{Multiscale phases and relative misfit reduction to validation misfit at initial phase iteration with number of iterations per phase}
 \centering
 \begin{tabular}{l c c c}
 \hline
 Stage & Minimum period (s) & Misfit reduction (\%) & Number of Iterations  \\
 \hline
  1 & 50  &  45 & 53   \\
  2 & 35  & 23 & 37  \\
  3 & 28  & 8 & 75  \\
  4 & 22  & 5 & 80  \\
  5 & 18  & 4 & 40 \\
 \hline
 \multicolumn{2}{l}{}
 \end{tabular}
 \label{misfit_table}
 \end{table}
%

%
\begin{figure}
    \centering
    \includegraphics[width=1.0\linewidth]{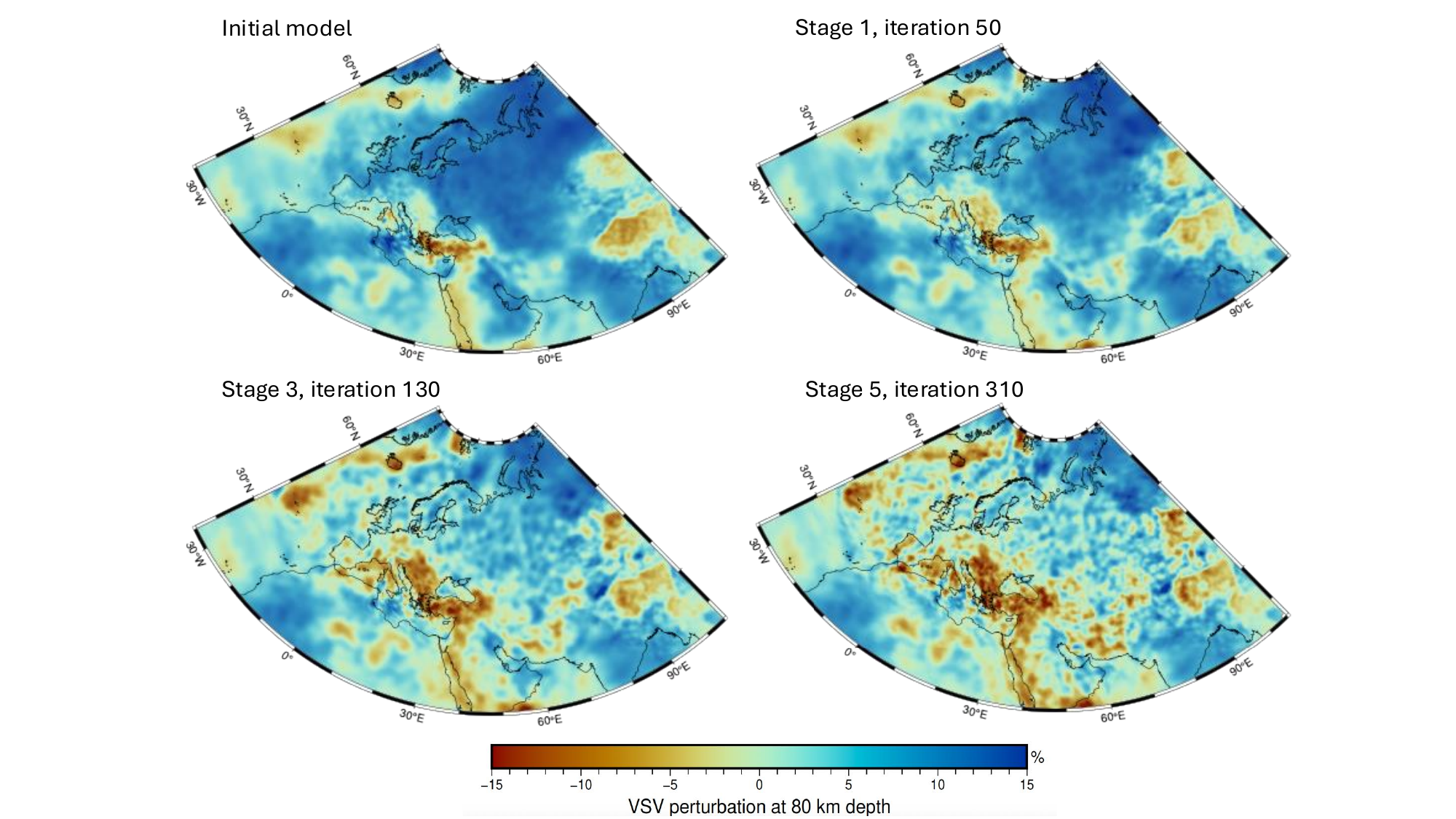}
    \caption{Evolution of perturbative changes to the model over the course of the inversion project. We can see a jump in detail with Stage 3 that corresponds to period band 28 s - 90 s.}
    \label{fig:model_development}
\end{figure}
%

%
\begin{figure}
    \centering
    \includegraphics[width=1.0\linewidth]{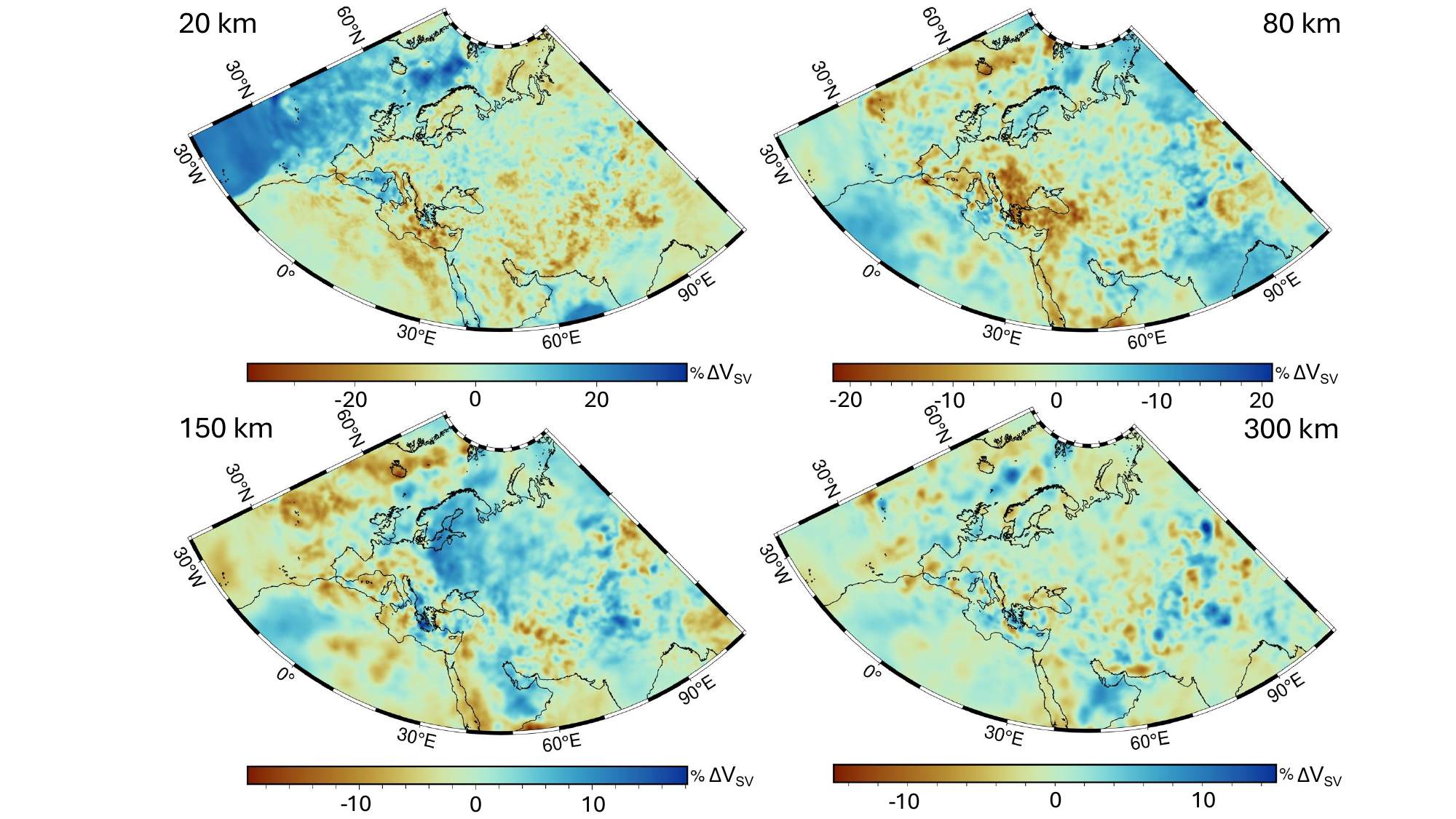}
    \caption{Depth slices for the vertical shear wave speed field of EUWA}
    \label{fig:EUWA_depths}
\end{figure}
%

%
\begin{figure}
    \centering
    \includegraphics[width=1.0\linewidth]{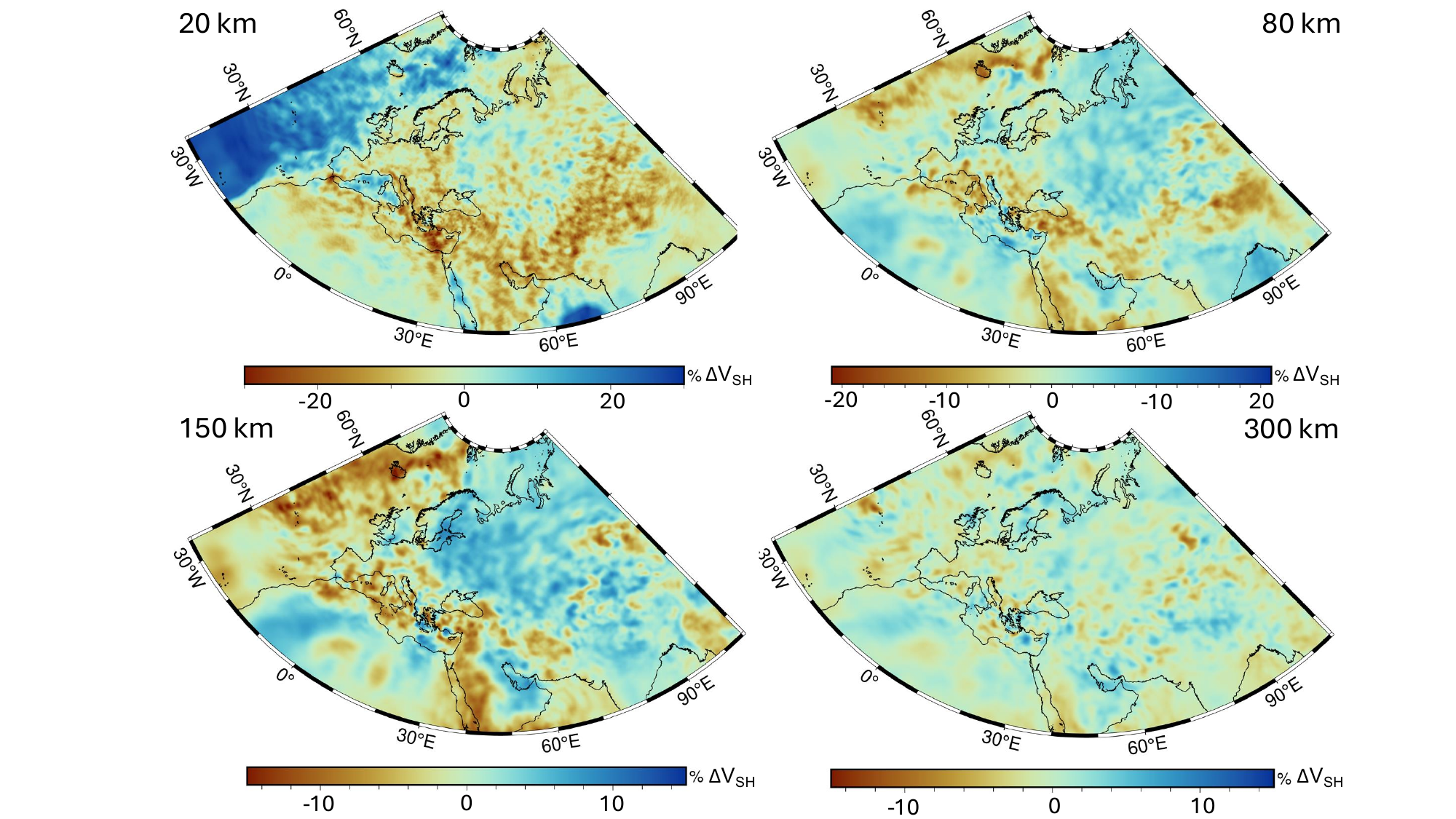}
    \caption{Depth slices for the horizontal shear wave speed field of EUWA}
    \label{fig:EUWA_depths}
\end{figure}
%


\section{Number of elements}

\begin{figure}
    \centering
    \includegraphics[width=1\linewidth]{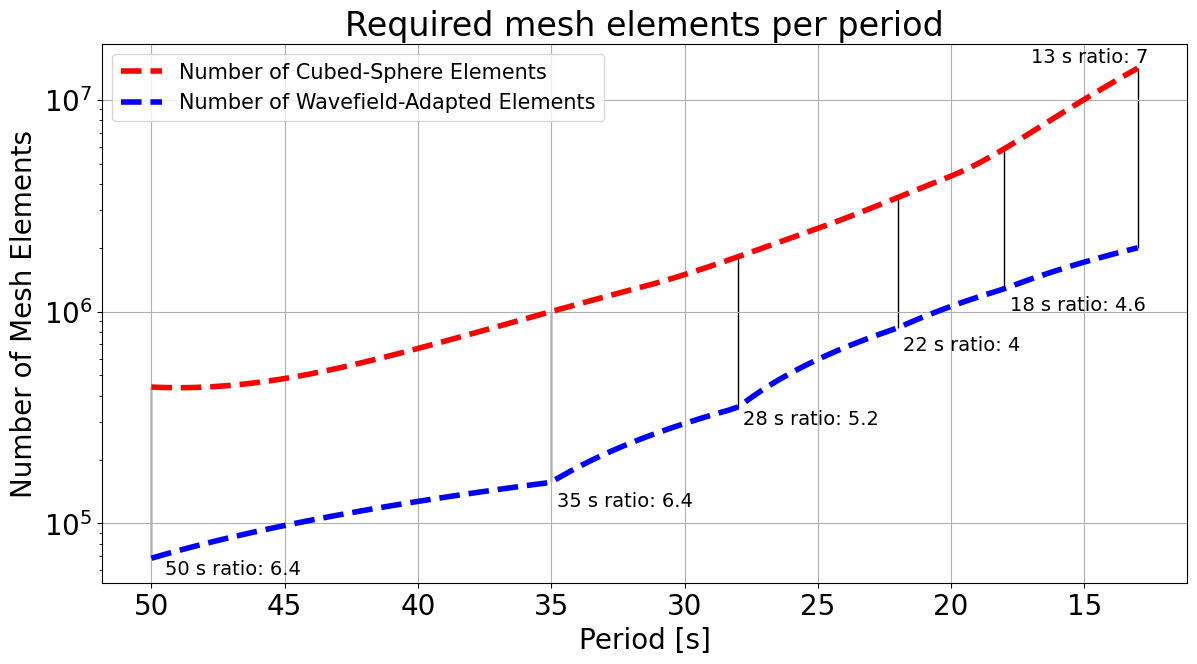}
    \caption{Number of elements needed for a mesh to sufficiently discretize the wavefield at a given minimum period with two elements per wavelength.  An element size difference ratio between 4 and 7 is maintained at different minimum periods. At 18 s, a cubed-sphere mesh has 5$\,$875$\,$000 elements, while an equivalent wavefield-adapted mesh has 1$\,$280$\,$000 elements for this domain.}
    \label{compute costs}
\end{figure}

